\documentclass[graybox]{svmult}

\usepackage{mathptmx}        
\usepackage{helvet}          
\usepackage{courier}         

\usepackage{makeidx}         
\usepackage{graphicx}        
\usepackage{multicol}        
\usepackage[bottom]{footmisc}

\usepackage{url}             

\makeindex             

\begin{document}




\mainmatter
%
%
%

{ 

\title*{Turbulent Boundary Layer in a 3-Element High-Lift Wing: Coherent Structures Identification}
\titlerunning{Turbul. Bound. Layer 3-Elem. High-Lift Wing: Coherent Struct. Identif.}
\author{R. Montal\`a \and B. Eiximeno \and A. Mir\'o \and O. Lehmkuhl \and I. Rodr\'iguez}
\institute{R. Montal\`a \and I. Rodr\'iguez \at Universitat Polit\`ecnica de Catalunya, \email{{ricard.montala, ivette.rodriguez}@upc.edu}
\and B. Eiximeno \and A. Mir\'o \and O. Lehmkuhl \at Barcelona Supercomputing Center, \email{{benet.eiximeno, arnau.mirojane, oriol.lehmkuhl}@bsc.es}}
\maketitle

\section{Introduction}
The reduction of fuel consumption in airplanes has always stand as a very appealing topic for the aeronautical industry. Economical savings and a lower environmental impact are the main benefits that companies aim to reach with it. In this sense, the breakthroughs in computational sciences, as well as the major advances in flow analysis, have allowed to disentangle the complexity of turbulence and gain insight into drag physics. However, these techniques has been usually limited to the analysis of canonical fluid flows, such as plane turbulent boundary layers or channel flows.

In the present work, we propose to address the physics of a more complex configuration, closer to the geometry of a real wing. Specifically, computational predictions on the 30P30N three-element high-lift wing are conducted, which has been used as a reference case for the AIAA Workshop on Benchmark Problems for Airframe Noise Computations (BANC). Consequently, many numerical contributions can be found on the literature\cite{Bodart, Jin, Zhang}, as well as experimental studies\cite{Klausmeyer,Murayama, Pascioni}. However, most of these works are focused on the aeroacoustic noise and, more precisely, on the flow mechanisms occurring at the slat cove.

Therefore, this work aims to extend the knowledge on high-lift wings and assess, not only the noise generation, but also identify the main causes of aerodynamic drag. The present results are the partial conclusions of a much larger comprehensive analysis which targets to identify the most energy-containing coherent structures and frequencies, and relate them to the Reynolds stresses and drag. The considered Reynolds number and angle of attack are $Re_c = 750,000$ and $AoA = 9^\circ$, respectively.

\section{Numerical method}

In the present work, a wall-resolved large-eddy simulation (WRLES) is conducted on the 30P30N high-lift wing geometry employing the finite-element code Alya. In Alya, the convective operator of the equations is approximated by a low-dissipation scheme\cite{Lehmkuhl}. In this approach, the energy, momentum and angular momentum are preserved at the discrete level, providing enhanced results. The set of LES equations is time-advanced using an energy conserving third-order Runge-Kutta explicit method combined with an eigenvalue based time-step estimator. A non-incremental fractional-step method is used to stabilise the pressure. For the turbulence modelling, the Vreman\cite{Vreman} eddy-viscosity model is considered.

\section{Results}

The 2D geometry of the airfoil is positioned within a circular computational mesh that extends a radius of $R=10C$ along the x-y plane and $L_z=0.1C$ along the z direction, as recommended by Lockard and Choudhari (2009)\cite{Lockard}, being $C$ the stowed chord. This direction is assumed periodic and is discretized using 128 planes. As the inflow boundary condition, a uniform velocity profile is applied, whereas zero-gradients are imposed at the outflow regions. A total of 58 million grid points are employed and a structured-like inflation layer is considered around the airfoil, allowing to achieve a non-dimensional near wall distance of $\Delta y^{+} \approx 1$. Here, a no-slip boundary condition is prescribed. Along the remaining directions, the maximum near wall distance is $\Delta x^{+}_{max} = 80$ and $\Delta z^{+}_{max} = 50$.

\begin{figure}[t]
\centering
\includegraphics[width=115mm]{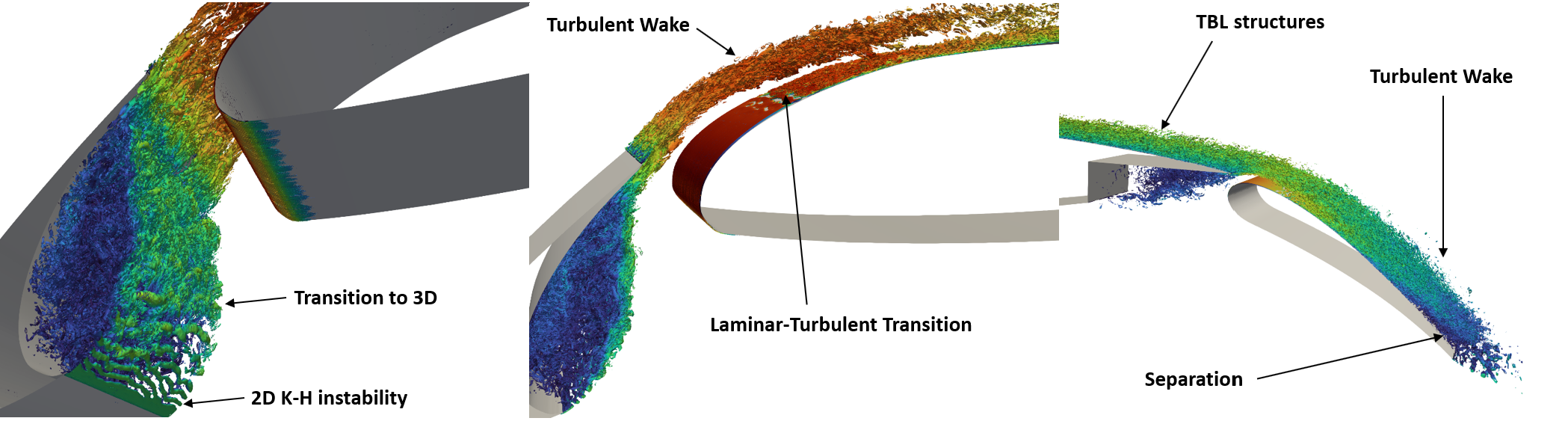}
\caption{Vortical structures represented by Q-isosurfaces}
\label{Montala_Q}
\end{figure}

 The predicted Q-criterion isosurfaces are depicted in figure \ref{Montala_Q}. As can be observed, the fluid flow past a three-element airfoil exhibits a wide variety of phenomena: Shear layers, a laminar-turbulent transition, bounded and wake turbulence, among others flow physics. The pressure and skin friction coefficients are plotted in figure \ref{Montala_CpCf}, along with selected results from the literature at comparable Reynolds numbers. Despite those experiments were performed at different flow conditions ($Re_c$ and $AoA$), results show reasonably good agreement. The computed lift coefficient is equal to $C_L=3.0879$, which lies in the range of values obtained by Murayama et al.\cite{Murayama} ($C_L=3.2428$) and Pascioni et al.\cite{Pascioni} ($C_L=3.0559$); whereas the drag coefficient has a value of $C_D=0.0885$ and no experimental results are available. The present work is focused on the bounded turbulence along the suction side of the main element.

\begin{figure}[t]
\centering
\includegraphics[width=100mm]{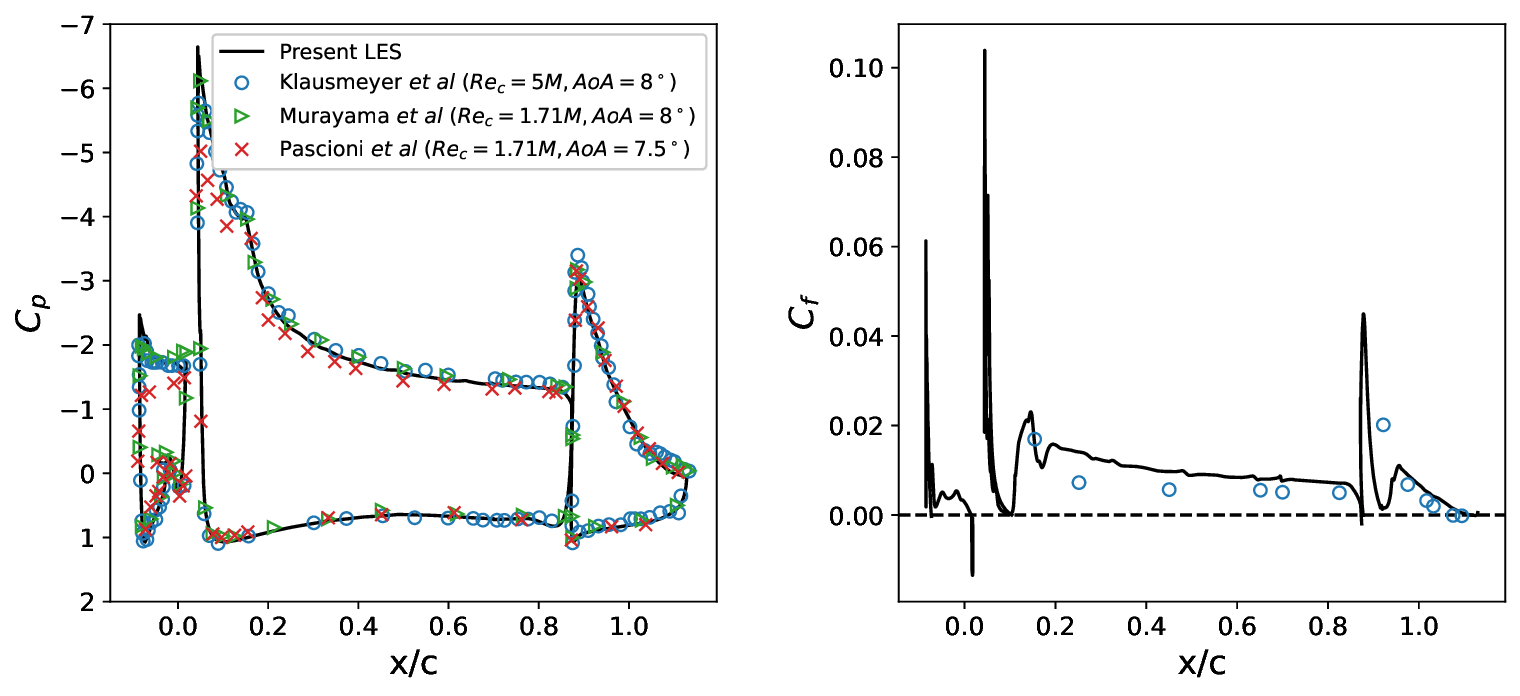}
\caption{Pressure $C_{p}$ (left) and skin-friction $C_{f}$ (right) coefficients. Skin friction represented only along the suction side. Present LES (black solid line) compared to the literature: Klausmeyer et al. (blue circles)\cite{Klausmeyer}, Murayama et al. (green triangles)\cite{Murayama} and Pascioni et al. (red crosses)\cite{Pascioni}.}
\label{Montala_CpCf}
\end{figure}

\subsection{Boundary Layer Development at Main Suction Side}

The turbulent boundary layer (TBL) development can be observed in figure \ref{Montala_BL}. In this figure, the boundary layer thickness, the friction Reynolds number, the Clauser pressure-gradient parameter and the shape factor are shown. It can be detected that, despite being an adverse pressure gradient (APG), the Clauser parameter is relatively small. This yields to a moderate growth of the boundary layer. In fact, the shape factor tends to $H=1.40$, which is the typical value observed in zero pressure gradient (ZPG) TBLs. Also notice that, near the trailing edge, $H$ suddenly decreases to a negative value due to the acceleration experimented by the fluid across the main-flap gap, leading to a local favorable pressure gradient (FPG).

\begin{figure}[t]
\centering
\includegraphics[width=100mm]{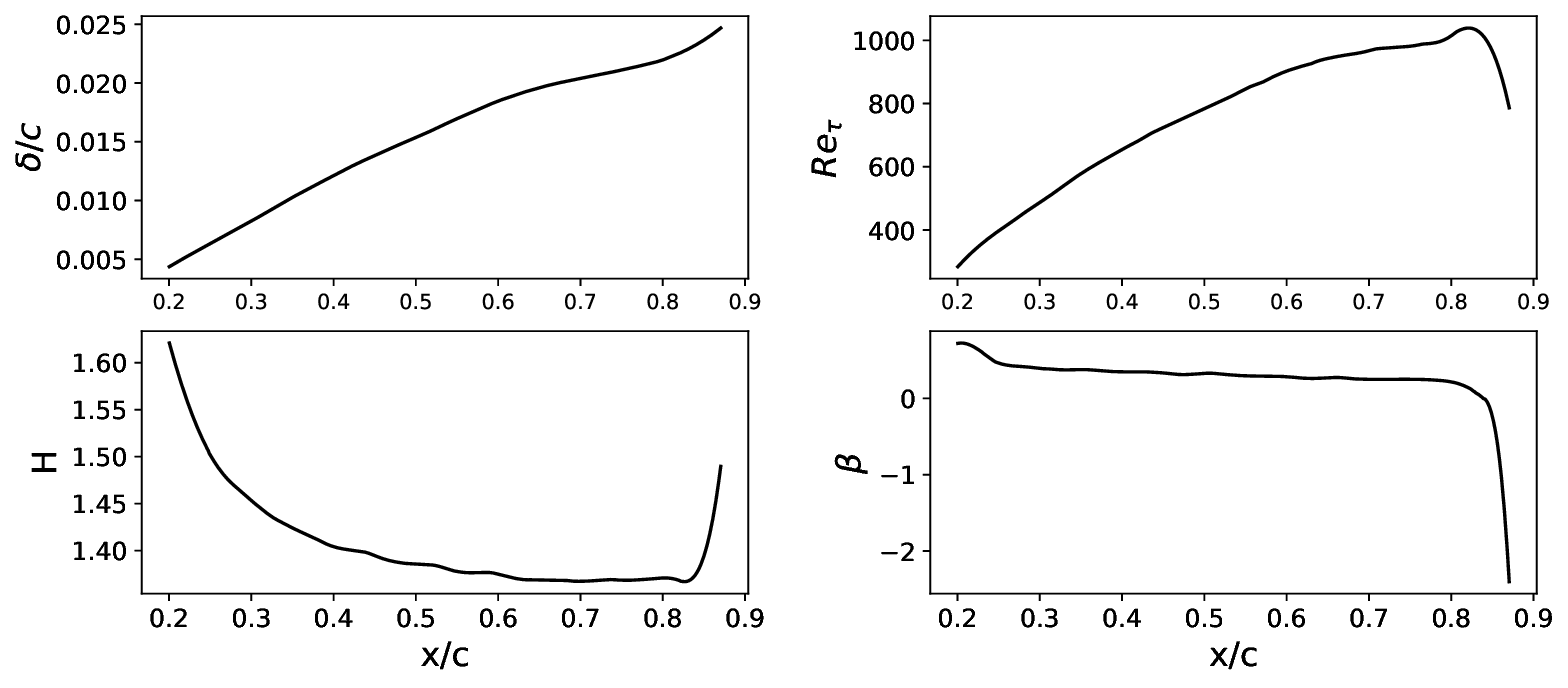}
\caption{Boundary layer thickness $\delta$ (top-left), friction Reynolds number $Re_{\tau}$ (top-left), shape factor $H$ (bottom-left) and Clauser pressure-gradient parameter $\beta$ (bottom-right) on the suction side of the wing.}
\label{Montala_BL}
\end{figure}

In figure \ref{Montala_RST}, the streamwise and wall-normal Reynolds stresses are plotted at three different locations and compared with the computations of other authors. Please refer to table \ref{Montala_RSTtab} for further information about these works. Notice that the historical effects at $x/c=0.25$ are visible, most likely caused by the turbulent wake coming from the slat and the laminar-turbulent transition occurring nearby (see fig. \ref{Montala_Q}). On the other hand, a better agreement with the literature for both the magnitude and the shape of the stresses is observed at $x/c=0.4$ and $0.7$. In fact, as commented before, results resemble those of a ZPG TBL rather than a APG TBL. Thus, the inner $\overline{u'u'}$ peak can be easily identified, while the outer one is roughly present, which is a typical feature from APG TBL.

\begin{figure}[t]
\centering
\includegraphics[width=110mm]{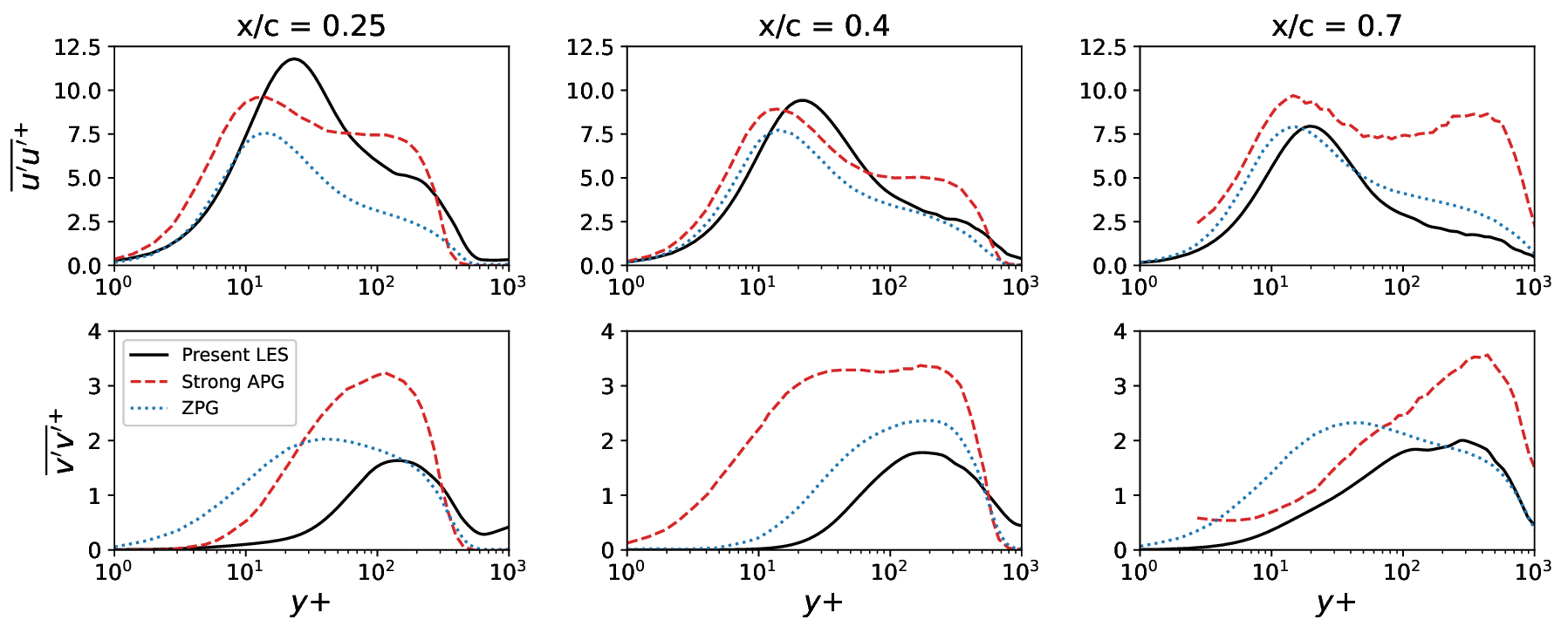}
\caption{Streamwise $\overline{u'u'}^{+}$ (top) and wall-normal $\overline{v'v'}^{+}$ (bottom) Reynolds Stresses at three different locations: $x/c = 0.25, 0.4$ and $0.7$ (from left to right). Results compared with the literature at matched friction Reynolds number $Re_{\tau}$: Present LES (black solid line); Strong APG (red dashed line) and ZPG (blue dotted line).}
\label{Montala_RST}
\end{figure}

\begin{table}
\caption{Summary of the literature employed in figure \ref{Montala_RST}.}
\label{Montala_RSTtab}
\begin{tabular}{p{2cm}p{2cm}p{1.5cm}p{1.5cm}p{1.5cm}p{1.5cm}p{1cm}}
\hline\noalign{\smallskip}
Location & Case & $Re_{\tau}$ & $Re_{\theta}$ & H & $\beta$ & Ref.  \\
\noalign{\smallskip}\svhline\noalign{\smallskip}
 & Present LES &  397 & 1,026 & 1.50 & 0.46 & -\\
$x/c=0.25$ & Strong APG & 373 & 1,722 & 1.74 & 4.10 & \cite{Sanmiguel}\\
& ZPG & 492 & 1,421 & 1.43 & 0 & \cite{Schlatter}\\

& Present LES & 655 & 1,626 & 1.40 & 0.34 & -\\
$x/c=0.4$ & Strong APG & 671 & 2,877 & 1.58 & 2.00 & \cite{Vinuesa}\\
& ZPG & 671 & 2,001 & 1.41 & 0 & \cite{Schlatter}\\

& Present LES & 968 & 2,498 & 1.37 & 0.25 & -\\
$x/c=0.7$ & Strong APG & 1,070 & - & - & 2.4 & \cite{Sanmiguel} \\
& ZPG & 1,139  & 3,600 & - & 0 & \cite{Eitel} \\
\noalign{\smallskip}\hline\noalign{\smallskip}
\end{tabular}
\end{table}

\subsection{POD analysis}

In order to get more insight into the TBL structures present, a proper orthogonal decomposition (POD)\cite{Lumley} is conducted. For the study, a total number of 620 snapshots are taken every 250 simulation time steps. Only a small portion of the domain is considered for this analysis, comprising $x/C \in [0.55, 0.80]$, $y/C \in [0.04, 0.15]$ and the whole span-length.

Figure \ref{Montala_POD} shows the reconstruction of the fluid field employing the first one hundred modes, as well as the cumulative turbulent kinetic energy (TKE) distribution. The latter indicates a wide energy spread along high order modes, requiring more than 360 modes to recover the 80\% of the TKE. This may be explained due to the great amount of similar coherent structures present in a TBL, and that energy is homogeneously distributed, making the POD algorithm to fail in separating them efficiently. Despite this large energy spread, modes are capturing the most prominent structures, i.e. the boundary layer streaks.

In figure \ref{Montala_Phi}, the isocountours of the POD spatial modes $\phi$ along a transverse plane located at $x/c=0.675$ are displayed. It is interesting to highlight how the most energetic $\phi_u$ structures are located at the height of $\overline{u'u'}_{peak}$, whereas the $[\phi_v, \phi_w]_{mag}$ structures are located nearby the region of $\overline{v'v'}_{peak}$. Also notice that for $\phi_u$, there are also secondary structures around the primary ones that might be related to the outer $\overline{u'u'}$ hump. 

\begin{figure}[t]
\centering
\includegraphics[width=60mm]{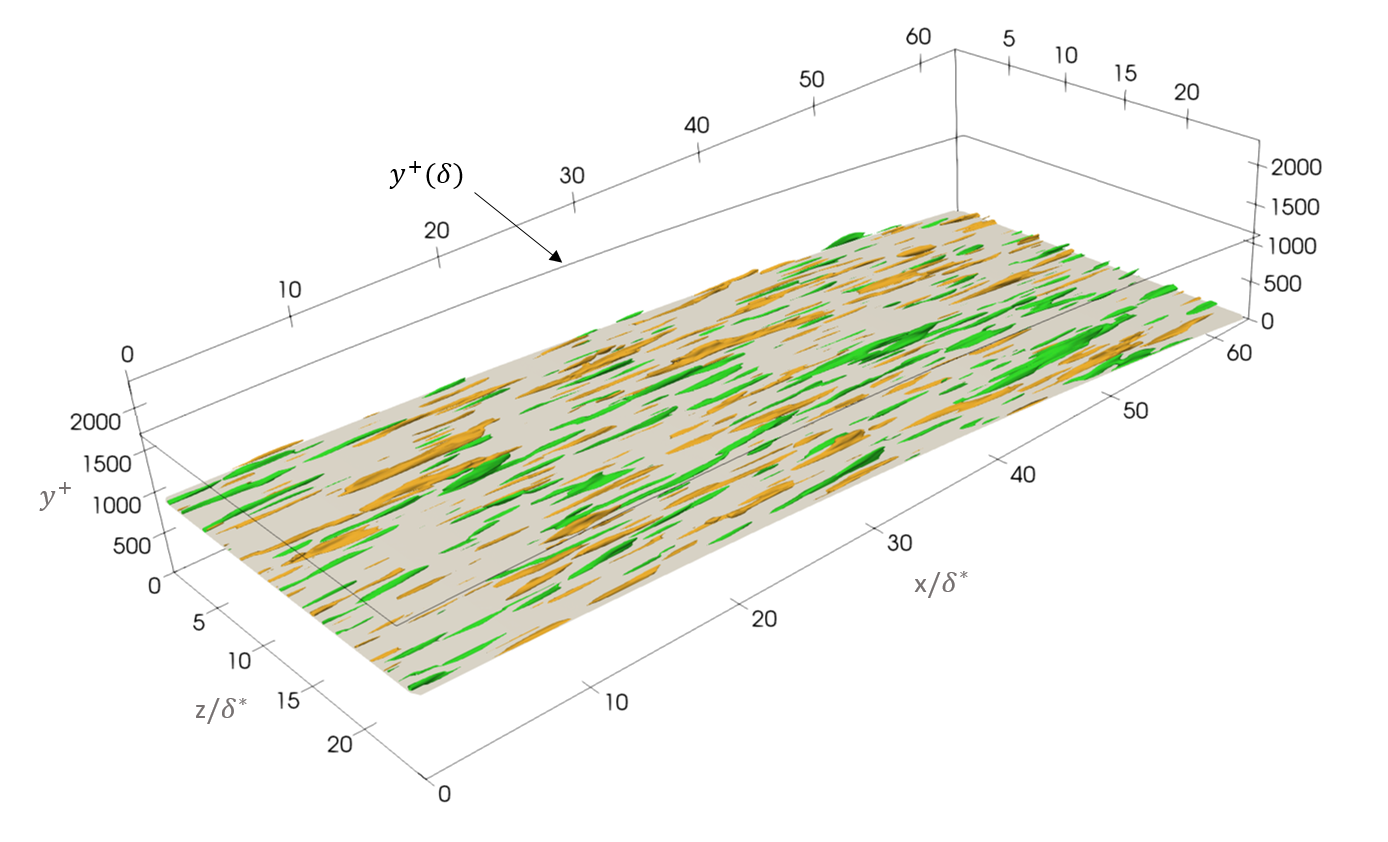}
\includegraphics[width=45mm]{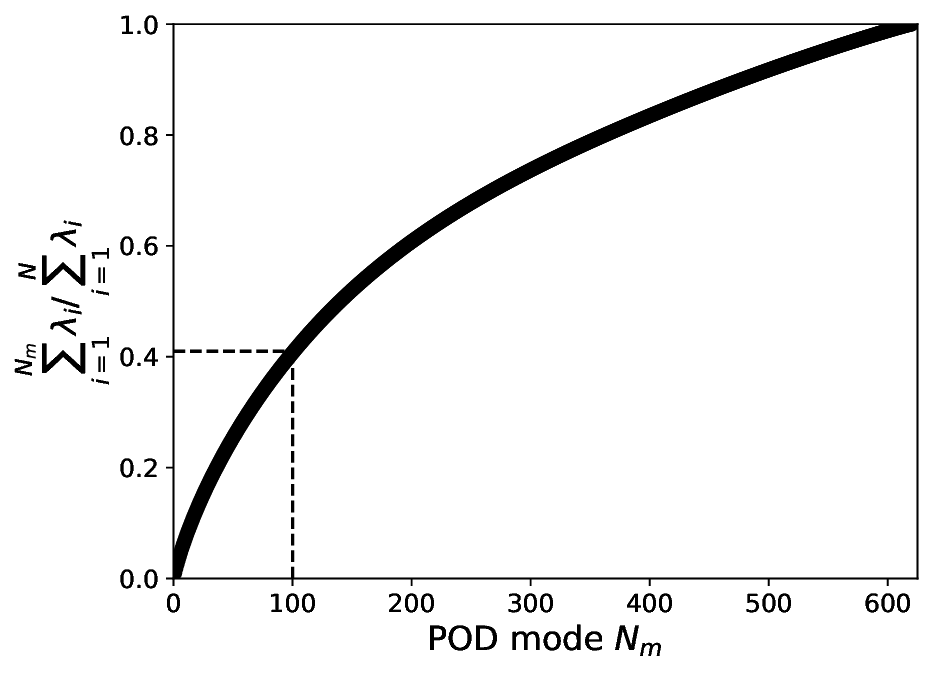}
\caption{Streamwise velocity $u$ reconstruction using 100 POD modes (left) and cumulative sum of POD eigenvalues i.e. TKE (right).}
\label{Montala_POD}
\end{figure}

\begin{figure}[t]
\centering
\includegraphics[width=115mm]{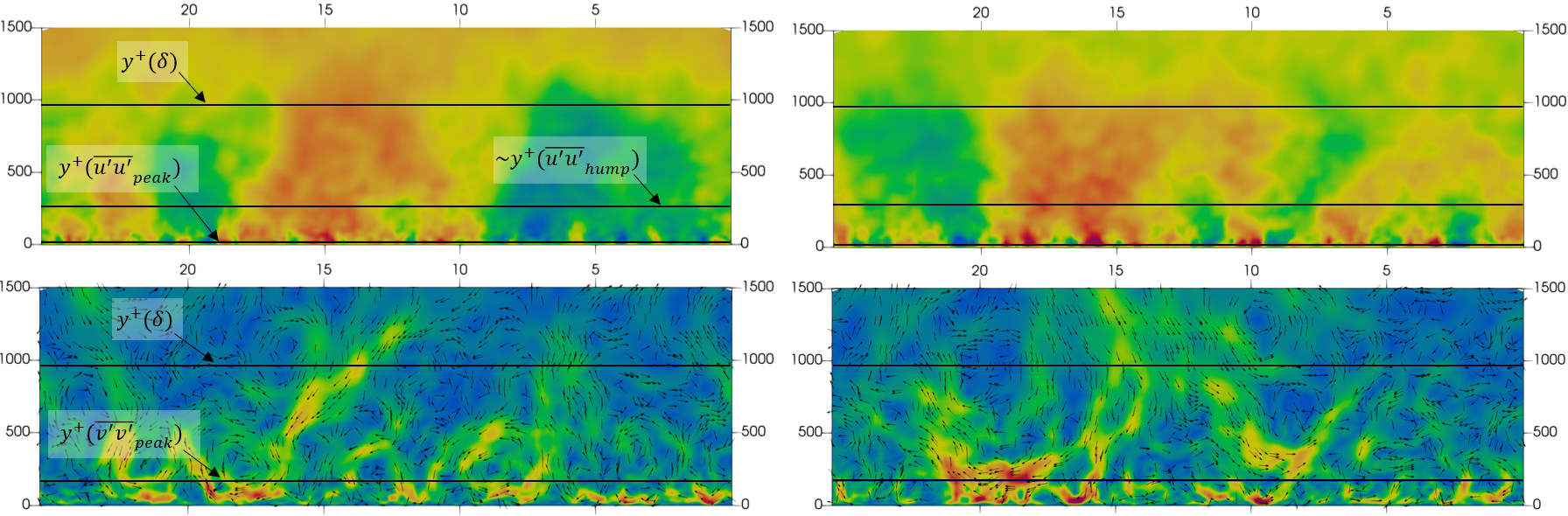}
\caption{$\phi_u$ (top) and $[0,\phi_v,\phi_w]_{mag}$ (bottom) isocontours at the transversal plane located at $x/c=0.675$ for modes 1 and 3 (left and right, respectively).}
\label{Montala_Phi}
\end{figure}

\section{Conclusions}
A wall-resolved large-eddy simulation of the fluid around a 30P30N airfoil is performed at a Reynolds number of $Re_c=750,000$ and an angle of attack of $AoA=9^\circ$. Results are validated with experiments carried out by other authors and further assessed in more detail along the suction side of the main element. Specifically, the boundary layer development is analysed, exhibiting a behaviour close to a ZPG TBL. Thus, the growth of the boundary layer is not pronounced and the outer peak of the streamwise Reynolds stresses is practically not present, which is a typical footprint of a APG TBL. A POD analysis is applied to a small portion of the TBL. In spite of the large energy spread to high order modes, TBL streaks are identified; and the location of the most energetic structures are associated to the peaks in the Reynolds stresses.  

\section*{Acknowledgments}
This research has received financial support from the {\it Ministerio de Ciencia e Innovación} of Spain (PID2020-116937RB-C21 and PID2020-116937RB-C22). Furthermore, simulations were conducted with the assistance of the {\it Red Española de Supercomputación} (RES) and the EuroHPC JU, who granted us computational resources at the HPC facilities of {\it MareNostrum IV}, at Barcelona Supercomputing Center (IM-2022-3-0005), and {\it Vega}, at Institute of Information Science (EHPC-REG-2022R01-030), respectively. Ricard Montalà work is funded by a FI-SDUR grant (2022 FISDU 00066) from the {\it Departament de Recerca i Universitats de la Generalitat de Catalunya}. Oriol Lehmkuhl has been partially supported by a {\it Ramon y Cajal} postdoctoral contract (RYC2018-025949-I). The authors also acknowledge the support of the {\it Departament de Recerca i Universitats de la Generalitat de Catalunya} for funding the research groups Large-scale Computational Fluid Dynamics (2021 SGR 00902) and Turbulence and Aerodynamics Research Group (2021 SGR 01051).


}

%



\end{document}